\documentclass[twocolumn,preprintnumbers, amssymb,amsmath,aps,floatfix,nofootinbib,superscriptaddress,showpacs]{revtex4-1}

\usepackage{epsfig}
\usepackage{bm}
\usepackage{amssymb}
\usepackage{amsmath}
\usepackage{color}
\usepackage{subfigure}
\usepackage[colorlinks,
            linkcolor=blue,
            anchorcolor=black,
            citecolor=blue
            ]{hyperref}
            
\newcommand{\be}{\begin{equation}}
\newcommand{\ee}{\end{equation}}
\newcommand{\bea}{\begin{eqnarray}}
\newcommand{\eea}{\end{eqnarray}}

\begin{document}
\title{Exploring the Collective Phenomenon at the Electron-Ion Collider}

\author{Yu Shi} 
\affiliation{Key Laboratory of Quark and Lepton Physics (MOE) and Institute of Particle Physics, Central China Normal University, Wuhan 430079, China}

\author{Lei Wang}
\affiliation{Key Laboratory of Quark and Lepton Physics (MOE) and Institute of Particle Physics, Central China Normal University, Wuhan 430079, China}

\author{Shu-Yi Wei}   \email{swei@ectstar.eu}
\affiliation{European Centre for Theoretical Studies in Nuclear Physics and Related Areas (ECT*)
and Fondazione Bruno Kessler, Strada delle Tabarelle 286, I-38123 Villazzano (TN), Italy}

\author{Bo-Wen Xiao}   \email{xiaobowen@cuhk.edu.cn}
\affiliation{School of Science and Engineering, The Chinese University of Hong Kong, Shenzhen 518172, China}

\author{Liang Zheng} \email{zhengliang@cug.edu.cn}
\affiliation{School of Mathematics and Physics,
China University of Geosciences (Wuhan),
Wuhan 430074, China }

\begin{abstract}
Based on rare fluctuations in strong interactions, we argue that there is a strong physical resemblance between the high multiplicity events in photo-nuclear collisions and those in $pA$ collisions, in which interesting long range collective phenomena are discovered. This indicates that the collectivity can also be studied in certain kinematic region of the upcoming Electron-Ion Collider (EIC) where the incoming virtual photon has a sufficiently long lifetime. Using a model in the Color Glass Condensate formalism, we first show that the initial state interactions can explain the recent ATLAS azimuthal correlation results measured in the photo-nuclear collisions, and then we provide quantitative predictions for the long range correlations in $eA$ collisions in the EIC regime. With the unprecedented precision and the ability to change the size of the collisional system, the high luminosity EIC will open a new window to explore the physical mechanism responsible for the collective phenomenon.
\end{abstract}

\maketitle

\section{Introduction} 

Collective phenomenon seems to be ubiquitous and is observed almost everywhere in high energy hadron-hadron collisions. Observations of the non-trivial azimuthal angle correlations (also known as flow harmonics) in heavy ion collisions, i.e. nucleus-nucleus collisions, have informed us a lot of interesting physics regarding the collective behavior and other physical properties of quark gluon plasma. Moreover, greatly to our surprise, when only high multiplicity events are selected, unexpected collectivity can also be found in small collisional systems such as proton-nucleus and proton-proton collisions. There has also been tremendous amount of undisputed evidence\cite{Khachatryan:2010gv, CMS:2012qk, Abelev:2012ola, Aad:2012gla, Adare:2013piz, Adare:2014keg, Khachatryan:2015waa, PHENIX:2018lia} which suggests the existence of the long range collective phenomenon in small systems in the last decade in both RHIC and the LHC. In addition, sizable signals of collectivity have been found not only for soft and light hadrons but also for heavy flavor mesons\cite{CMS:2018xac, Acharya:2017tfn, Sirunyan:2018toe, CMS:2019isc} in small systems.  

Central to a lot of experimental and phenomenological studies on the collectivity are the physics origin and quantitative interpretation of the long range correlation in small systems. The collectivity is quantitatively defined as the Fourier coefficients of the azimuthal angular correlation of the measured particle $v_n\equiv \langle \cos n (\phi-\Psi_n) \rangle$, where $\phi$ is the azimuthal angle of the measured particle and $\Psi_n$ is the reference angle (i.e., the reaction plane angle). Conventionally, $v_n$ is also known as the $n$-th flow harmonics, since the relativistic hydrodynamics framework can quantitatively and successfully explain\cite{arXiv:1304.3044,arXiv:1304.3403,1306.3439,arXiv:1307.4379,arXiv:1307.5060,arXiv:1312.4565,arXiv:1405.3605,Habich:2014jna, arXiv:1409.2160, arXiv:1609.02590,arXiv:1701.07145,arXiv:1801.00271} the collective behavior of soft light hadrons measured at both RHIC and the LHC. In this framework, the underlying physics degrees of freedom becomes relativistic fluids, since the number of produced particles after initial collisions are usually assumed to be sufficiently large in high multiplicity events. As a result, the collective behavior of final state particles is interpreted as the final state energy anisotropy of the evolved fluid with certain initial spatial anisotropy. Also, there have been several other alternative interpretations based on particles scattering models and kinetic theories, see examples in Refs.~\cite{Lin:2003jy,arXiv:1803.02072, Li:2018leh, Kurkela:2018qeb}. Additional final state analysis\cite{Du:2018wsj} also indicates that final state effects can only generate a fraction of the elliptic flow $v_2$ for heavy mesons measured at the LHC \cite{CMS:2018xac, Acharya:2017tfn, Sirunyan:2018toe}. 

Another competitive explanation of the observed collectivity in small systems comes from initial state interactions\cite{Armesto:2006bv, Dumitru:2008wn, Gavin:2008ev, Dumitru:2010mv, Dumitru:2010iy, Kovner:2010xk, Kovchegov:2012nd, Dusling:2012iga, Dumitru:2014dra, Dumitru:2014yza, Dumitru:2014vka, Lappi:2015vha, Schenke:2015aqa, Lappi:2015vta,McLerran:2016snu, Kovner:2016jfp, Iancu:2017fzn, Dusling:2017dqg, Dusling:2017aot, Fukushima:2017mko, Kovchegov:2018jun, Boer:2018vdi, Mace:2018vwq, Mace:2018yvl,Kovchegov:2013ewa,Altinoluk:2018ogz,Kovner:2018fxj,Kovner:2017ssr,Kovner:2018vec, Davy:2018hsl, Zhang:2019dth, Zhang:2020ayy} in the so-called Color Glass Condensate (CGC) framework, which is widely viewed as the effective theory of Quantum Chromodynamics (QCD) when the gluon density is high. In CGC, dense gluons in a high energy hadron typically carry finite amount of transverse momentum at the order of the saturation momentum $Q_s$. For example, the amount of transverse momentum broadening that a high energy quark receives after traversing a dense nuclear target is roughly $Q_{sA}^2$, with $Q_{sA}$ the corresponding saturation momentum of the nuclear target. The multiple interactions between the quark probe and the dense gluon target can be described by a color dipole in the coordinate space. Now suppose one considers the interactions between two initially uncorrelated quarks and a target nucleus. The transverse momentum broadening of these two quarks then can be characterized by two independent dipole scattering amplitudes, which contain no correlations. Interestingly, as shown in Refs~\cite{Mace:2018vwq, Mace:2018yvl, Davy:2018hsl}, these two dipoles can also be converted into a quadrupole\cite{Blaizot:2004wv, Dominguez:2008aa, Dominguez:2011wm, Dominguez:2012ad} during the interaction and non-trivial two particle azimuthal correlations can arise as the $1/N_c^2$ correction to the independent dipole scattering amplitudes. Using the extension of the CGC model from Refs~\cite{Mace:2018vwq, Mace:2018yvl}, not only can one explain the sizable $v_2$ for $J/\psi$ and open charm\cite{Zhang:2019dth} measured in $pPb$ collisions, but also make a further prediction\cite{Zhang:2020ayy} for the open bottom meson, which is confirmed by the recent CMS observation\cite{CMS:2019isc}.

Recently, there have been some more interesting experimental results regarding the two particle correlations in $e^+e^-$ collisions at LEP\cite{Badea:2019vey} and in deep inelastic $ep$ scattering at HERA\cite{ZEUS:2019jya}. First, the experimental effort based on the analysis of the archived data collected by the ALEPH detector at LEP so far does not find significant long-range correlations in high multiplicity $e^+e^-$ collisions. Second, the ZEUS collaboration measured the two particle azimuthal angle correlations in high multiplicity $ep$ collisions with virtuality $Q^2 >5 \text{GeV}^2$, and finds that the measured correlations are dominated by minijets contributions while the genuine collective phenomenon is not observed. On the other hand, recent ATLAS analysis\cite{ATLAS:2019gsn,Aad:2021yhy} of the photo-nuclear ultra-peripheral ($PbPb$) collisions (UPC) indicates the persistence of collective phenomenon in $\gamma A$ collisions with the strength of correlations comparable to those measured in proton-proton and proton-lead collisions in similar multiplicity ranges.

The objective of this paper is to explore the possibilities of observing collectivity at the upcoming Electron-Ion Collider (EIC)\cite{Boer:2011fh, Accardi:2012qut, Aidala:2020mzt}. Recently, it has been announced that the cutting-edge high-luminosity EIC will be built at the Brookhaven National Laboratory in the near future. Based on the above-mentioned experimental observations and theoretical arguments, we believe that the planned EIC is in a unique position to study the collectivity in small collisional systems and it can help us unravel the corresponding underlying mechanism. EIC offers us both $ep$ and $eA$ collisions with different values of virtuality which provide us additional handles to change initial conditions for the target and the size ($\sim 1/Q$) of the collisional system. 

\section{Collectivity in $\gamma^\ast A$ collisions } 
Let us try to understand the above seemingly mixed signals from these three experimental results\cite{Badea:2019vey,ZEUS:2019jya, ATLAS:2019gsn,Aad:2021yhy} with photons involved. 
This scattering can be viewed as the collision between a virtual photon with virtuality $Q^2$ and the target nucleus. Photons, especially low-$Q^2$ ones, can have a very rich QCD structure. In the field theory language, a photon state can be schematically decomposed as follows
\be
|\gamma \rangle = |\gamma_0 \rangle + \sum_{m, n} |m\, q \bar q +n\, g \rangle + \sum_{\rho, \omega, \cdots}  |V \rangle +\cdots, 
\ee
where $|\gamma_0 \rangle$ represents a point like photon which knocks out a quark from the target hadron in leading order DIS. In the large $x_{\text{B}} \equiv \frac{Q^2}{s}$ regime, the dominant contribution is described by the point like photon state with the size of order $1/Q$. $s$ is the center-of-mass energy square of the $\gamma^* A$ system.

\begin{figure}[tbp]
\begin{center}
\includegraphics[width=7.3cm]{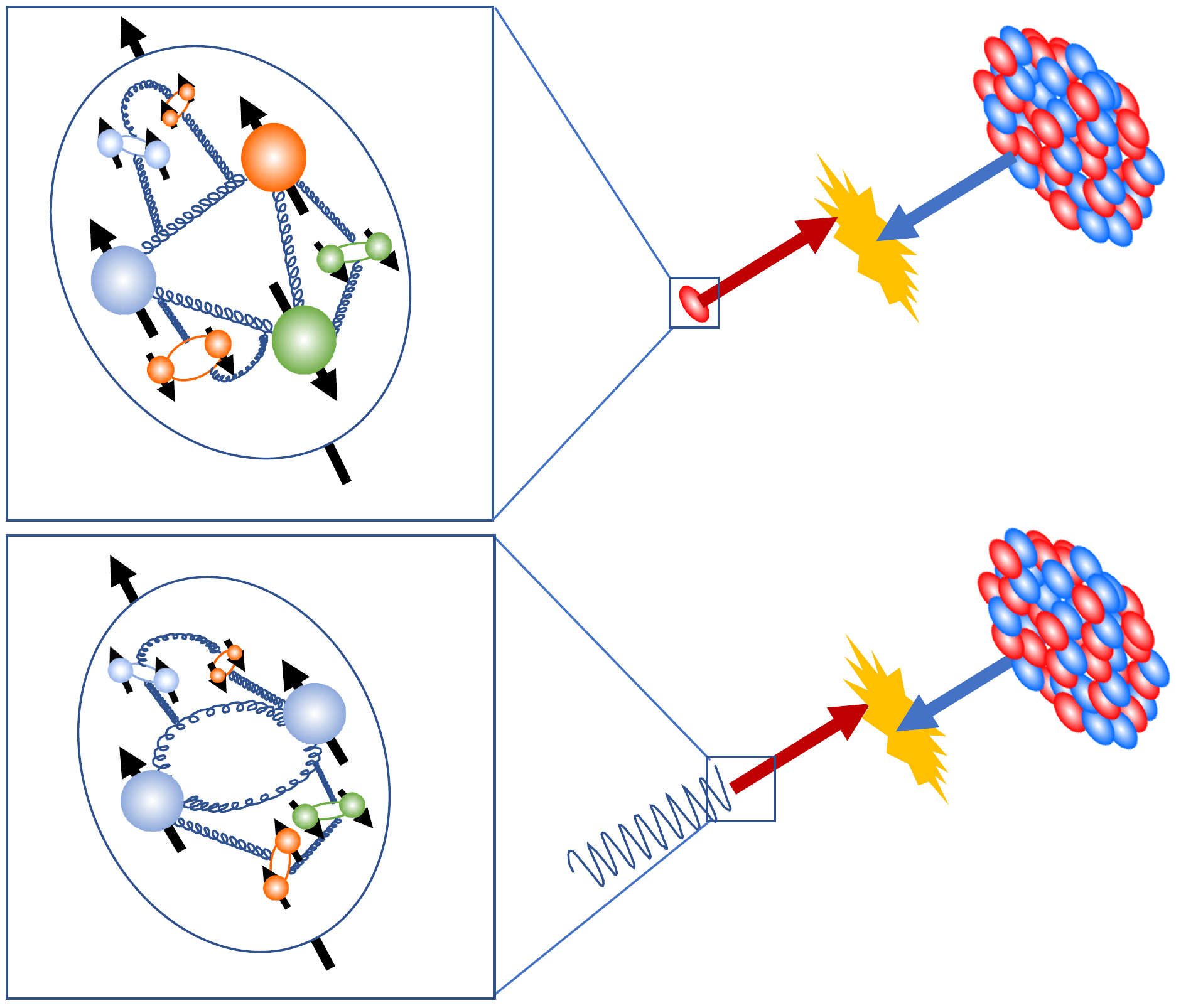}
\end{center}
\caption[*]{The cartoon illustrations of high multiplicity events in $pA$ and $\gamma^\ast A$ collisions where important QCD fluctuations of many active partons can be seen as we zoom in.}
\label{cartoon}
\end{figure}

More interesting parts of the photon structure can arise due to fluctuations when $x_\text{B}$ is sufficiently small. For example, a virtual photon can fluctuate into a pair of quark-antiquark (i.e., a color dipole), which is perturbatively calculable in high $Q^2$ regime. In the so-called Mueller's dipole frame\cite{Mueller:1989st, Mueller:1993rr, Kovchegov:2012mbw}, one can find that the lifetime of the virtual photon fluctuation becomes much longer than the time of its interaction with the target hadron, when $x_\text{B} \ll 1/(2M R)$ with $M$ the nucleon mass and $R$ the size of the target hadron. In general, a photon can fluctuate into an arbitrary number of $q\bar q$ pairs and gluons and eventually emerge as a ``color cloud". In other words, it can have non-trivial partonic substructure\cite{Schuler:1996fc, Nisius:1999cv} as well as rare fluctuation\cite{Mueller:2014fba, Liou:2016mfr}. Furthermore, in the low $Q^2$ regime, a photon state may also be decomposed into a set of vector meson states including $\rho\,, \omega\, , \phi$ and heavy quarkonia in the vector meson dominance model\cite{Sakurai:1960ju}. In high multiplicity events, due to the rare fluctuation with sufficiently long lifetime, the incoming low-$Q^2$ virtual photon can also be viewed as a hadron (i.e., a vector meson) with a large number of collinear partons as illustrated in Fig.~\ref{cartoon}. 

In this sense, in light of the strong resemblance between the virtual photon and the hadron in hadronic reactions, we believe that the high multiplicity events in $eA$ DIS in the low $x_\text{B}$ and low $Q^2$ regime is physically equivalent to those in $pA$ collisions, which is independent of the underlying interpretation of the collective phenomenon. Therefore, as argued in \cite{ATLAS:2019gsn,Aad:2021yhy}, there should be collective phenomena in photo-nuclear collisions as well. As to DIS with large $Q^2$\cite{ZEUS:2019jya} and $e^+e^-$ annihilations\cite{Badea:2019vey}, the high multiplicity events are dominated by the productions of mini-jets, which in principle contains little long range correlation. Here, we focus on the partonic contents of the photon wavefunction, since they provide a convenient description of the interactions in high energy collisions.


Experimentally, the discovery of the collective phenomenon strongly relies on the trigger selection of the rare events with extremely high multiplicities. From the theoretical perspective, the high multiplicity event first requires the participance of many active partons in the scattering. In particular, for small systems such as $pPb$ collisions, this implies that one should consider the rare fluctuation which creates a large number of active partons inside the proton wave-function. In DIS, similar many-body partonic structure can also arise from the wavefunction of virtual photons due to the QCD fluctuation. As to the target nucleus side, in addition to the possible large number of participating nucleons in the scattering, one can also expect stronger parton density in many of those nucleons which leads to larger overall saturation momentum. 

Based on the above assumptions, we can follow the CGC model developed in Refs~\cite{Mace:2018vwq, Mace:2018yvl, Davy:2018hsl, Zhang:2019dth, Zhang:2020ayy} and compute the corresponding azimuthal angular correlation in $\gamma^\ast A$ collisions by treating the virtual photon as a hadron with a lifetime longer than the time of interaction. For convenience, our calculation is carried out in the Breit frame. First, we use the following ansatz for the Wigner distribution to describe the distribution of partons inside the virtual photon projectile
\be
w(x, b_\perp, k_\perp) = f_{p/\gamma}(x) \frac{1}{\pi^2} e^{-b_\perp^2/B_p -k_\perp^2/\Delta^2},
\ee
where $f_{p/\gamma}(x)$ stands for the collinear parton distribution in the photon projectile with the longitudinal momentum fraction $x$, and the impact parameter $b_\perp$ and the initial transverse momentum $k_\perp$ of the parton are assumed to be of the Gaussian form with the corresponding variances $B_p$ and $\Delta^2$, respectively. Roughly speaking, $B_p$ characterizes the spread of partons in transverse coordinate space, while $\Delta$ gives the typical transverse momentum of the parton. For proton, one can take $B_p =6 \text{GeV}^{-2}$ which is related to the proton size. As to the virtual photon, since the size of QCD fluctuation is usually confined within the scale $1/\Lambda_{\text{QCD}}$, we set $B_p \sim \text{min}[1/Q^2\, , 1/\Lambda_{\text{QCD}}^2]$ based on the uncertainty principle. 

Second, the parton density in the projectile (e.g., proton or $\gamma^\ast$) is assumed to be much lower than that in the target hadron (e.g., heavy nucleus), thus the so-called dilute-dense factorization can be safely applied to the calculation. In the formalism, partons from the projectile traverse the background gluon fields of the target hadron, and then they get produced in the final state with typical transverse momentum of the order of $Q_s$. The above physical picture of the multiple scattering with the dense gluon fields in the target hadron essentially can be captured by the Wilson line ($U(x_\perp)$) in the eikonal approximation. After squaring the amplitude, one finds that the partonic process can be written as a color dipole in the coordinate space. For example, the production of a quark can be described by 
\be
\langle D(x_\perp,\, y_\perp)\rangle = \frac{1}{N_c} \langle \text{Tr} [U(x_\perp) U^\dagger(y_\perp)] \rangle,
\ee 
where $x_\perp$ and $y_\perp$ stand for the transverse coordinates of the quark in the amplitude and complex conjugate amplitude, respectively. Here $\langle \cdots \rangle$ represents the average over the dense background gluon fields in the target hadron. For simplicity, we usually approximately write $D(x_\perp, y_\perp) = \exp\left(-\frac{Q_s^2 r_\perp^2}{4}\right)$ with $r_\perp \equiv x_\perp -y_\perp$. The exponential form of the dipole amplitude can be understood as the result of the sum over arbitrary number of gluon exchanges with the target. It is then straightforward to see that the Fourier transform of the dipole amplitude yields a typical transverse momentum of $Q_s$ due to the multiple scattering with the target hadron for the final state produced quark. For an incoming gluon, one can simply replace the above quark dipole with a gluon dipole defined by the Wilson line in the adjoint representation. 

Last but not least, to illustrate the rise of the angular correlation in the CGC formalism, one can consider the production of two initially un-correlated quarks\footnote{The two quarks are picked from many active partons inside the photon wavefunction, therefore they are assumed to be un-correlated in both color and momentum. Similarly, other channels such as quark-gluon and gluon-gluon correlations have also been taken into account in this calculation.} in the dense gluon background fields of the target hadron, and find that the correlation appears as the higher order $N_c$ corrections in the resulting background average of two dipole amplitudes which reads
\begin{eqnarray}
&&\left.\langle D\left(b_1+\frac{r_1}{2},b_1-\frac{r_1}{2}\right)D\left(b_2+\frac{r_2}{2},b_2-\frac{r_2}{2}\right)\rangle \right|_{\text{up to }\frac{1}{N_c^2}}\nonumber
\\
&=&e^{-\frac{Q_s^2}{4}(r_1^2+r_2^2)}\left[1+\frac{1}{N_c^2} Q(r_1,  b_1, r_2, b_2)  \right],
\end{eqnarray}
where the first term represents two un-correlated dipoles produced in the final state. The second term inside the square brackets, which is proportional to
\bea
&&Q(r_1, b_1, r_2, b_2) \notag \\
&=& \left(\frac{Q_s^2}{2}r_1\cdot r_2\right)^2 \int_0^1 d\xi \int_0^{\xi} d\eta
e^{\frac{\eta Q_s^2}{8} [(r_1-r_2)^2-4(b_1-b_2)^2]} \,\,\,
\eea
comes from the color transition between the dipole configuration and the quadrupole configuration. $r_1$ and $r_2$ represents the transverse sizes of these two dipoles, and $b_1$ and $b_2$ stand for their transverse locations. Similar as the calculation laid out in Refs~\cite{Mace:2018vwq, Mace:2018yvl, Davy:2018hsl, Zhang:2019dth, Zhang:2020ayy} for $pA$ collisions, the multi-particle spectra and correlations in high energy $\gamma^\ast A$ collisions then can also be obtained from the Fourier transform of the above dipole amplitudes, when we treat the incoming virtual photon as a hadron with many active partons in the high multiplicity events. Nevertheless, it is worth mentioning that the transverse size $\sim 1/Q$ of the incoming photon can vary significantly in contrast to the fixed size of the proton. In this model calculation, we have completely discarded the contribution of the jet-type correlation, which is presumably removed in the experimental analysis of the long range correlations. 


\begin{figure}[tbp]
\begin{center}
\includegraphics[width=7.8cm]{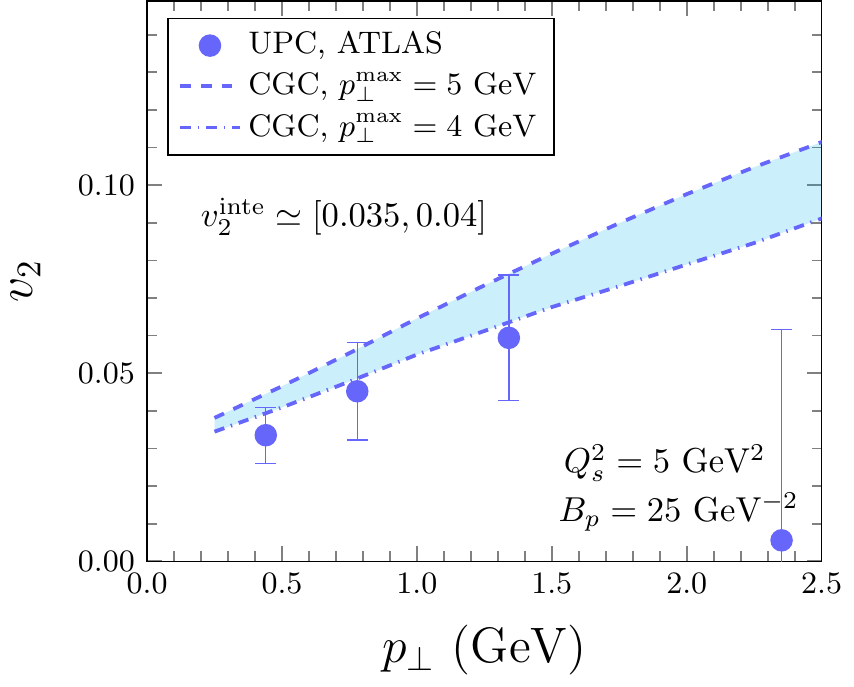}
\end{center}
\caption[*]{The comparison with the ATLAS photo-nuclear data\cite{ATLAS:2019gsn,Aad:2021yhy} and the resulting $v_2$ from the CGC model calculation by using $\Delta=0.5 \text{GeV}$ and $Q_s^2 =5 \, \text{GeV}^2$ which are the same as the parameters used in Refs.~\cite{Zhang:2019dth, Zhang:2020ayy}.}
\label{upc}
\end{figure}

\begin{figure}[tbp]
\begin{center}
\includegraphics[width=7.8cm]{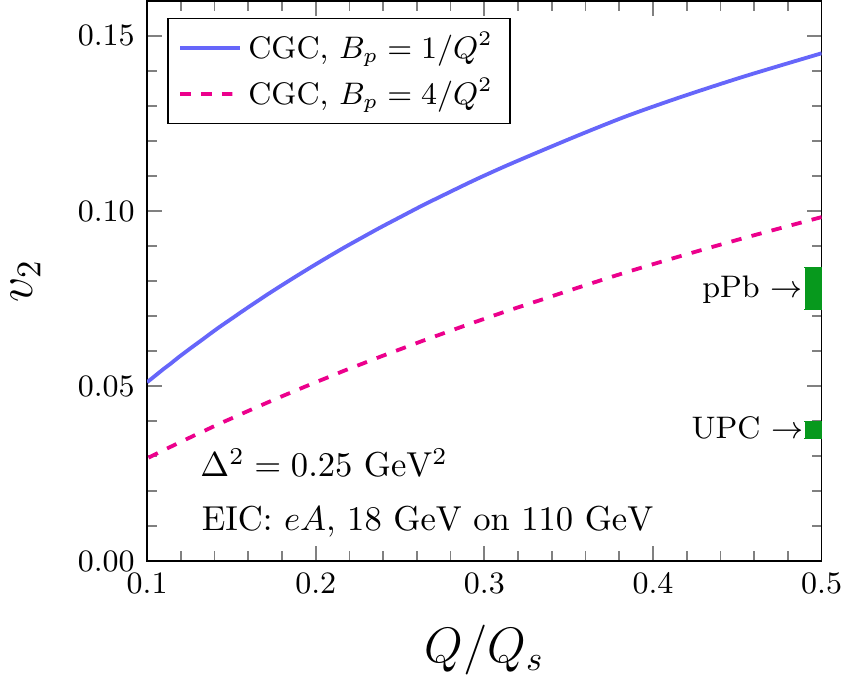}
\end{center}
\caption[*]{The prediction for integrated $v_2$ in the EIC regime. In this calculation, we fix the photon momentum fraction $y\equiv p\cdot q/p\cdot l = 0.9$. The theory curves are obtained by setting $Q_s^2=4$ GeV$^2$ and varying $Q$ between $0.2$ and $1$ GeV.}
\label{eic}
\end{figure}

In the two-particle correlation method, $v_2$ is defined as,
$
v_2 (p_\perp^a) \equiv v_{2,2} (p_\perp^a,p_\perp^b) / \sqrt{v_{2,2} (p_\perp^b,p_\perp^b)},
$
where, $v_{2,2} (p_\perp^a, p_\perp^b) \equiv \langle e^{i2(\phi_{a} - \phi_{b})}\rangle $ is the second Fourier harmonic of the differential two-particle spectrum with $p_\perp^a$ and $p_\perp^b$ representing different $p_\perp$ ranges for the trigger and associate particles, respectively.
In Fig.~\ref{upc}, we show the resulting two particle correlations $v_2$ for two different values of the maximum integrated transverse momentum $p_\perp^{\text{max}}$ as the function of hadron transverse momentum $p_\perp$ in photo-nuclear reactions in the above CGC model, and find them in agreement with the recent ATLAS data. Our results for the integrated $v_2$ (i.e., $v_2^{\text{inte}}$) is also in line with the ATLAS data. In this reaction, the typical virtuality ($Q$) of the incoming photon is usually of the order of $30\, \text{MeV}$\cite{Krauss:1997vr, Baur:2001jj} which is much smaller than $\Lambda_{\text{QCD}}$. However, the extent of the QCD fluctuation usually does not exceed the size $1/\Lambda_{\text{QCD}}$ due to the color confinement, and thus $B_p$ is set to be $25 \, \text{GeV}^{-2}$ in this special case. Although the integrated $v_2$ only weakly depends on the cut $p_\perp^{\text{max}}$\cite{Dusling:2017dqg, Dusling:2017aot}, the differential $v_2$ is also sensitive to the choice of $p_\perp^{\text{max}}$ when hadron fragmentation functions are used. Besides, it is important to note that the current CGC model employed here is only applicable\cite{Zhang:2019dth, Zhang:2020ayy} in the low $p_\perp$ regime. 

In Fig.~\ref{eic}, assuming $ \Lambda_{\text{QCD}} \leq Q < 1 \,\text{GeV}$ at EIC and setting $B_p = 1/Q^2$ (or $4/Q^2$), the predictions of the integrated $v_2$ in the regime of future EIC are shown as the function of $Q/Q_s$. This plot indicates that sizable collectivity comparable to that in UPC and $pPb$ collisions at the LHC is expected at EIC from the CGC perspective. By varying the virtuality ($Q$) of the incoming photon, we can study the system size dependence of the initial state interactions as well. Also, we notice that events with $Q^2$ as low as $0.045\, \text{GeV}^2$ were measured at HERA\cite{Aaron:2009aa}. As $Q$ increases with fixed $Q_s$, the system size decreases and the typical spatial distance between the trigger particle and the reference particle also shrinks, thus these two particles are more likely to scatter with the same color domain of the size $1/Q_s$ in the nuclear target. Since the correlation generated in the CGC model usually emerges within a color domain\cite{Lappi:2015vta, Dusling:2017dqg, Dusling:2017aot}, it is then natural to expect that $v_2$ increases with increasing $Q/Q_s$ ratio. Nevertheless, as previously argued, our model is only applicable in the low-$Q^2$ region where the ratio $Q/Q_s$ is small. In addition, $v_2$ only weakly depends on the value of $\Delta$, which is assumed to be much smaller than $Q_s$ (This is equivalent to say that the parton density in the target nucleus is much higher than that in the incoming photon). Therefore, the resulting $v_2$ at the EIC is only sensitive to the dimensionless quantity $Q/Q_s$.

\section{Discussion and Summary} 

Let us make some further comments on several interesting aspects of the collective phenomenon and the resulting impact on the future EIC research efforts. 

First, the high luminosity EIC will offer an unprecedented opportunity to study the collective behavior of high multiplicity events. In particular, we argue that the system size and collisional energy can be adjusted by selecting high multiplicity events with different values of photon virtuality $Q^2$ and energy fraction $y$, respectively. Compared to the UPC data from ATLAS\cite{ATLAS:2019gsn,Aad:2021yhy} with the integrated luminosity of $1.73$ nb$^{-1}$, the exploration of collectivity in high multiplicity events should be more statistically favored at the future EIC with the planned integrated luminosity 10 fb$^{-1}$/year. On the other hand, it appears that this study in $ep$ collisions could be more challenging even in the EIC era depending on the event statistics and the underlying mechanism. In $ep$ collisions, the strength of the saturation effect may not be sufficient in the context of CGC interpretation, and the number of high multiplicity events is also a limiting factor. Nevertheless, it is certainly of great importance to compare the study in EIC to the analysis of HERA data, which may cast light on the origin of collectivity and rare QCD fluctuations.

Furthermore, from the point of view of Monte Carlo simulation, spatial and momentum correlations between the interacting partons can arise through the nonlinear QCD evolution and the multiple scattering in the dipole model~\cite{Avsar:2010rf, Domine:2018myf}. In addition, it is interesting to note that a new event generator\cite{Bierlich:2019wld} for $\gamma^\ast A$ collisions based on Mueller's dipole evolution\cite{Mueller:1993rr} is currently under development. This allow us to study the initial partonic geometries of proton and nucleus related to the collective phenomenon. Sophisticated implementations of these multiple parton interaction contributions in the PYTHIA/Angantyr\cite{Bierlich:2016smv} can provide simulations for the parton spatial distributions and their fluctuations in $eA$ and $pA$ collisions. 

Finally yet importantly, the initial state interpretation in terms of the CGC model may not the only explanation for the collectivity in $\gamma^\ast A$ collisions (if it is confirmed in the EIC or other experimental studies). The contribution of final state effects is also of great interest. Sizable initial eccentricities together with final state interactions such as hydrodynamics or other final state strong interactions imply that the similar collective phenomenon may arise in the high multiplicity DIS events.

In summary, we have analyzed high multiplicity events in DIS and argued that the collective phenomenon can also be explored at EIC based on the physical similarity between $pA$ and $\gamma^\ast A$ collisions in these events. In a simplified CGC model, we first show that initial state effect can describe the recent ATLAS data measured in the photo-nuclear ultra-peripheral $PbPb$ collisions, and we further make predictions for the two particle correlations at the planned EIC which can be studied in much more detail. Eventually, future efforts in this direction may lead us to a fundamental understanding of the origin of the collectivity in high energy collisions.

\begin{acknowledgments}
We thank Z.Y. Chen, A. Deshpande, Y. Hatta, W. Li, D. Perepelitsa, R. Venugopalan, N. Xu and F. Yuan for inspiring discussions and comments. This material is partly supported by the Natural Science Foundation of China (NSFC) under Grant Nos.~11575070 and 11905188. 
\end{acknowledgments}

\end{document}